\documentclass{sig-alternate}[12pt]

\begin{document}
%
% --- Author Metadata here ---
%\conferenceinfo{WOODSTOCK}{'97 El Paso, Texas USA}
%\CopyrightYear{2007} % Allows default copyright year (20XX) to be over-ridden - IF NEED BE.
%\crdata{0-12345-67-8/90/01}  % Allows default copyright data (0-89791-88-6/97/05) to be over-ridden - IF NEED BE.
% --- End of Author Metadata ---

\title{Improving an Hybrid Literary Book Recommendation System through Author Ranking}
 \numberofauthors{2} %  in this sample file, there are a *total*
% % of EIGHT authors. SIX appear on the 'first-page' (for formatting
% % reasons) and the remaining two appear in the \additionalauthors section.
% %
 \author{
% % You can go ahead and credit any number of authors here,
% % e.g. one 'row of three' or two rows (consisting of one row of three
% % and a second row of one, two or three).
% %
% % The command \alignauthor (no curly braces needed) should
% % precede each author name, affiliation/snail-mail address and
% % e-mail address. Additionally, tag each line of
% % affiliation/address with \affaddr, and tag the
% % e-mail address with \email.
% %
% 1st. author
\alignauthor
Paula Cristina Vaz\\
       \affaddr{INESC-ID/IST}\\
       \affaddr{Rual Alves Redol, 9}\\
       \affaddr{1000029 Lisbon, Portugal}\\
       \email{paula.vaz@inesc-id.pt}
% 2nd. author
\alignauthor
David Martins de Matos\\
       \affaddr{INESC-ID/IST}\\
       \affaddr{Rual Alves Redol, 9}\\
       \affaddr{1000029 Lisbon, Portugal}\\
       \email{david.matos@inesc-id.pt}cm/cmr6.pf
% 3rd. author
\and
\alignauthor
       Bruno Martins\\
       \affaddr{IST/INESC-ID}\\
       \affaddr{Avenida Professor Cavaco Silva}\\
       \affaddr{2780990 Porto Salvo, Portugal}\\
       \email{bruno.martins@ist.utl.pt}
% 4th. author
\alignauthor Pavel Calado\\
       \affaddr{IST/INESC-ID}\\
       \affaddr{Avenida Professor Cavaco Silva}\\
       \affaddr{2780990 Porto Salvo, Portugal}\\
       \email{pavel.calado@ist.utl.pt}
}

\date{30 July 1999}

\maketitle

\begin{abstract}
Literary reading is an important activity for individuals and choosing to read a book can be a long time commitment, making book choice an important task for book lovers and public library users. In this paper we present an hybrid recommendation system to help readers  decide which book to read next. We study book and author  recommendation in an hybrid recommendation setting and test our approach in the LitRec data set. Our hybrid book recommendation approach purposed combines two item-based collaborative filtering algorithms to predict books and authors that the user will like. Author predictions are expanded in to a book list that is subsequently aggregated with the former list generated through the initial collaborative recommender. Finally, the resulting book list is used to yield the top-n book recommendations. By means of various experiments, we demonstrate that author recommendation can improve overall book recommendation.
\end{abstract}

\category{H.3.3}{Information Storage and Retrieval}{Information filtering}
%A category including the fourth, optional field follows...
\category{H.3.7}{Digital Libraries}{System issues}

\terms{Algorithms, Experimentation}

\keywords{Hybrid Recommender, Book Recommendation, Author Recommendation}
~\\
\section{Introduction}
%Motivation
Literary reading is an important activity for individuals.
Public libraries make it possible to exercise this activity for free, by letting users borrow books for one or two weeks. In this context, choosing the right book to borrow becomes an important task, because it can save the reader unnecessary trips to the library to pick new books.

On-line recommendation systems have proved to be very useful helping, users through the suggestion of items that satisfy user needs or preferences. Good recommendations in a public library could improve reader's usability of the library.
%Problem
Libraries have limited shelf space, but still have enough books to make book selection difficult and time consuming. However, the number of books and users is not enough to successfully use the traditional collaborative techniques that rely on large amounts of data to detect patterns. Public library users have a limited number of books that can be borrowed each time they visit the library. In Portugal, most libraries set a limit of five books for a two week period. In this context, it is important that the top five recommendations have books preferred by the user.

%goals
This research work aims to $(i)$ assess whether item-based collaborative filtering (ICF) can be used to make good recommendations in an a public library, and $(ii)$ assess whether selecting books by author preferences can improve recommendations (a survey posted on \textit{Goodreads}\footnote{http://www.goodreads.com is a social network for book readers.} revealed that 78\% of the respondents choose the next book to read with basis on authorship).

%solution
In this paper, we purpose a weighted hybrid approach to recommend literary books, that can be used in the context of public libraries. Our approach combines two ICF algorithms to improve recommendations, where one recommends books (ICFB) and the other recommends authors (ICFA). Authors are used to improve the book top-n recommendations through a fusion approach.

\section{Background}
In this section we briefly explain and classify recommendation systems. We also describe the score aggregation functions used in our approach.
~\\

\subsection{Collaborative filtering}
Collaborative filtering is a technique used in recommendation where information is filtered using multiple sources. Literature on recommendation systems~\cite{adomavicius05} distinguishes between two main types of collaborative filters, namely user-based (UCF) and item-based (ICF).

UCF tries two find like-minded users to produce recommendations. Generally, this type of algorithms have poorer performance than ICF algorithms~\cite{sarwar01}.

ICF, popularized by \textit{Amazon.com}, searches for commonalities between items to make recommendations. Traditional ICF systems represent items as an N-dimensional vector of users, where N is the number of users in the system.  Each position of the vector contains the rating given by the user to the item. The algorithm computes an item similarity matrix, using an appropriate similarity function. The most common function used in these algorithms is the cosine similarity~\cite{sarwar01}. Finally, items similar to user preferred items are aggregated and ranked to generate recommendations.

\subsection{Aggregation functions}
To aggregate the items in ICF algorithms, several functions have been proposed. In this work we used the Reciprocal Rank Fusion (RRF)~\cite{cormack09} and Collaborative Filtering Preference Aggregation (CFPA)~\cite{beliakov11} to combine items recommended by two different ICF algorithms.

The RRF combines document rankings from multiple ranked lists. RRF sorts the documents according to a naive scoring formula. Given a set $D$ of documents and a set of rankings $R$, we compute the RRF score as shown in Equation~\ref{eq:rrf}.
\begin{equation}
\label{eq:rrf}
RRFscore (d \in D) = \displaystyle\sum_{r\in R} \frac{1}{60+r(d)}
\end{equation}
The CFPA  weights document similarity with the user rating for the document. Given a set $D$ of documents, a user $u$, and a set $S$ of similarities, we compute CFPA score for a given user as shown in equation~\ref{eq:cfpa}.
\begin{equation}
\label{eq:cfpa}
CFPAscore (u) = \displaystyle\sum_{i}R(u,d_{i}) \times \vec{d_{i}}
\end{equation}
In the formula, $R(u,d_{i})$ is the rating that user $u$ gave to document $d_{i}$ and $ \vec{d_{i}}$ is a column of the document similarity matrix.

Two combine the output of the two ICF algorithms we used the Weighted Arithmetic Mean (WAM)~\cite{beliakov11}, as show in equation~\ref{eq:wam}.
\begin{equation}
\label{eq:wam}
WAMscore (u) = \frac{\alpha~ICFA + (1-\alpha)~ICFB}{2}
\end{equation}

\subsection{Hybrid recommendation systems}
In addition to collaborative filtering techniques, recommendation system can also be content-, demographic-, and/or knowledge-based~\cite{burke07}. These techniques can be combined in a unique hybrid system. Hybrid systems combine two or more algorithms to improve recommendations, overcoming limitations of individual algorithms.
According to the classification given in~\cite{burke07}, in this work we will use a weighted system characterized by combining numerically the score of different recommendation components.
~\\

\subsection{Book and author recommenders}
There are countless book recommendation sites that can be found on the Internet. Of these, we highlight:  \textit{gnooks}\footnote{http://www.gnooks.com/} (can recommend books and authors. \textit{gnooks} includes a ``literature map" that graphically shows authors read together); \textit{Similar authors}\footnote{http://www.similarauthors.com/} (shows lists of authors similar to the author given user); \textit{BookLamp}\footnote{http://booklamp.org/} (defines the ``book DNA" where author information is included and is used to find similar books). In~\cite{kamps11}, the author investigates the effectiveness of author rankings in a library catalog to improve book retrieval. However, to the best of our knowledge, this is the first study attempting to improve book recommenders through author recommendation.

\section{Mean Reciprocal Rank}
To evaluate our approach we used the mean reciprocal rank (MRR). The MRR is a statistic used to evaluate the quality of top-n lists generated by retrieval processes. The MRR measures how far from the top appears the first good document and can be defined by Equation~\ref{eq:mrr}, where $p_i$ is the position in the top of the first good document.
\begin{equation}
MRR = \frac{\displaystyle\sum_{i=1}^{number~of~hits} \frac{1}{p_i}}{number~of~tops}
\label{eq:mrr}
\end{equation}
where a $hit$ is a well predicted document.

\section{The hybrid book recommender}
\begin{figure}
\centering{
\includegraphics[scale=0.28]{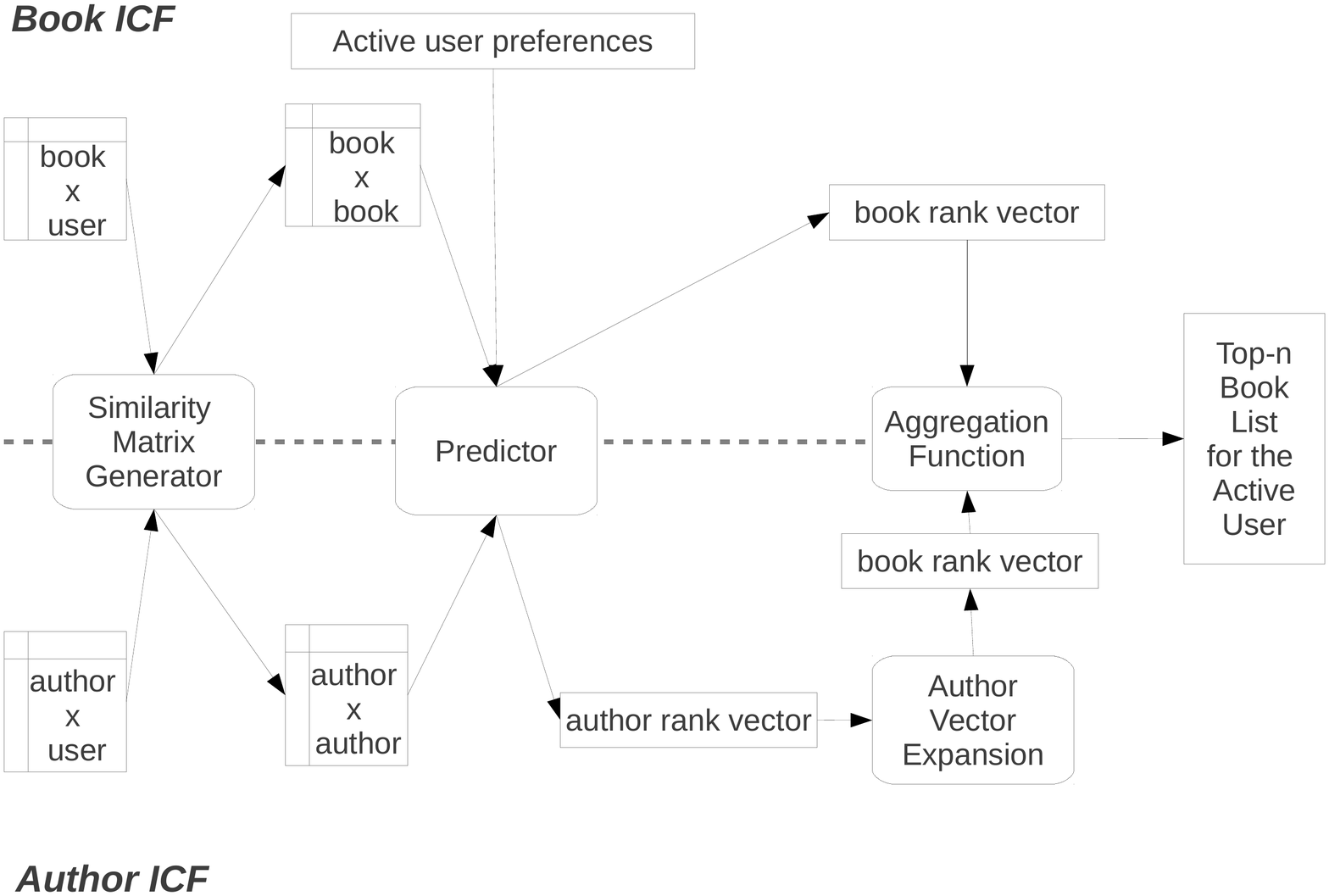}}
\caption{Hybrid book recommender algorithm.}
\label{fig:hba}
\end{figure}

This section describes the LitRec data set and discuss our findings. Our hybrid book recommender (HBR) algorithm, outlined in Figure~\ref{fig:hba}, is divided in three phases: the training or similarity matrix calculation, prediction, and aggregation phases.

\subsection{LitRec data set}
LitRec is a literary data set built for recommendation purposes. It combines documents from Project Gutenberg\footnote{http://www.gutenberg.org/} with ratings from  \textit{Goodreads}.

LitRec contains 38,591 ratings from 1,927 users and 3,710 documents.
The data set also contains book authors (1,627 different authors), the user location (1,029 different locations), the review date, and document content. The review date was used to sort and divide ratings in a train-test set of 90\%-10\%. 

\subsection{Similarity matrix calculation}
The HBR represents books as an N-dimensional vector of users where components contain the rating given by the user to the book. Ratings are integers in the 1 to 5 scale. Authors are represented as an N-dimensional vector of users where components contain the average of the ratings given by the user to the books written by the author.

The HBR generates book and author similarity matrices using book and author vectors and a similarity function. We made experiments using the cosine, the inverted Euclidean distance (Equation~\ref{eq:euclidean}), and co-occurrences.
\begin{equation}
ieuc(\vec{u},\vec{v}) = \frac{1}{\sqrt{(u_1 - v_1)^2 +...+(u_n - v_n)^2}}
\label{eq:euclidean}
\end{equation}
Co-occurrences between book $b_i$ and $b_j$ are calculated by adding one to $cell_{i,j}$ of the $book\times book$ similarity matrix every time $b_i$ and $b_j$ are preferred by the same user. Co-occurrences between authors follow the same approach. 

We also experimented using the cosine and Euclidean distance with 2nd-order vectors of co-occurrences~\cite{kulkarni05}. 2nd-order vectors represent items as an N-dimensional vector of items where each component contains the number of times the items co-occur.

\subsection{ICF prediction}
Book and author rank vectors (RV) (Figure~\ref{fig:hba}) are generated by the predictor using the similarity matrices and the active user\footnote{The active user is the user for which recommendations are being generated.} (AU) preference vector. To generate the book RV, the predictor selects the $book \times book$ matrix columns corresponding to the user's favorite books.

To calculate the author RV, the predictor counts the number of books that the AU preferred from each author. Then, it selects the $author \times author$ matrix columns corresponding to the user's favorite authors.

The retrieved columns are aggregated using RRFscore (Equation~\ref{eq:rrf}) and CFPAscore (Equation~\ref{eq:cfpa}). The main difference between the two scores is that CFPAscore weights the item columns with the user rating.

\begin{figure}
\includegraphics[scale=0.30]{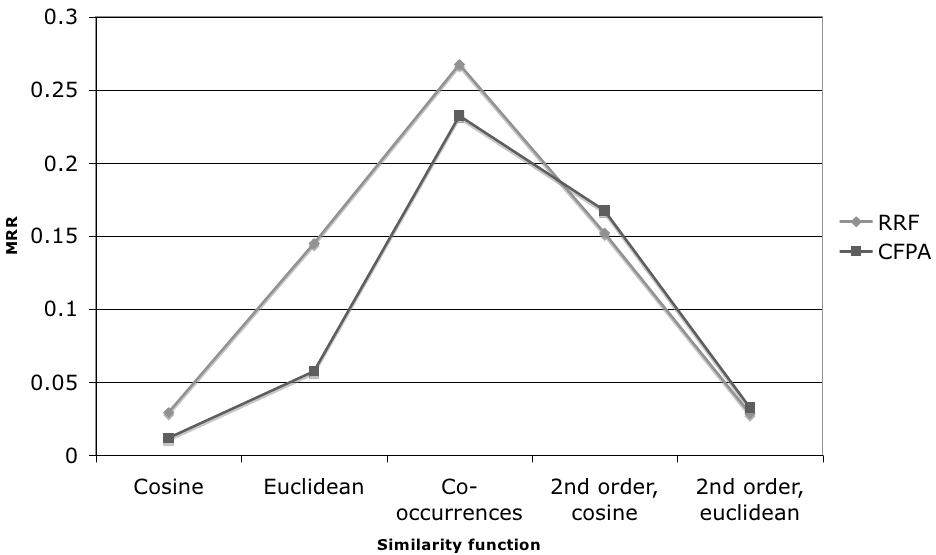}
\includegraphics[scale=0.30]{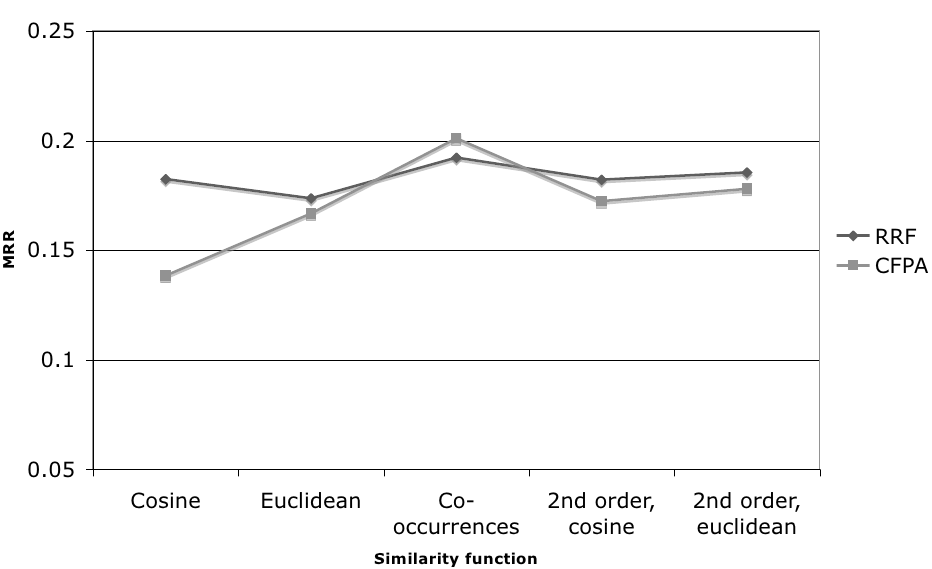}
\caption{Book (top) and author (bottom) MRR comparing different similarity measures.}
\label{fig:bookMRR}
\end{figure}

%\subsection{Similarity matrices evaluation}
The evaluation results using the MRR statistic are shown in Figure~\ref{fig:bookMRR}.
Evaluation shows that, for the LitRec data set, our algorithm produces better RV using co-occurrence matrices. However, author RV are more sensitive to user ratings than book RV.

As Figure~\ref{fig:bookMRR} outlines, the best book predictions were achieved with co-occurrence matrix and RRFscore aggregation,
whereas, the best author predictions were achieved with co-occurrence matrix and CFPAscore aggregation.

Compared to other data sets used for recommendation research, e.g., the Movilens\footnote{http://www.movielens.org}  data set, the number of ratings is much smaller. This generates very sparse rating matrices, leading to the poor performance of geometric similarity measures (cosine similarity and euclidean distance). In the remaining experiments, we used co-occurrence matrices to generate predictions.

\subsection{Author vector expansion}
After finding the authors similar to the AU favorite authors, the HBR expands the author RV to a book RV. The algorithm fills a book RV, assigning to each book its author rank weighted by the book popularity. Book popularity is measured by their frequency in the data set.

\begin{figure}
\includegraphics[scale=0.30]{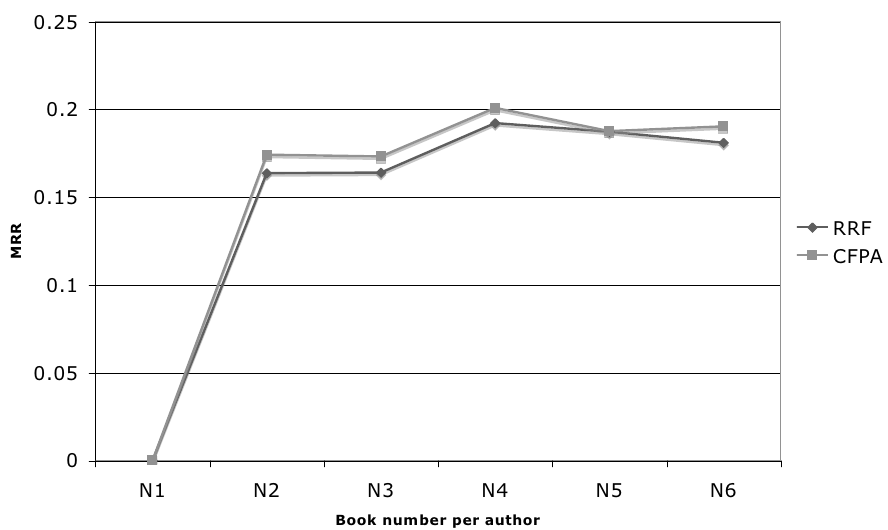}
\caption{Limit on book number per author evolution.}
\label{fig:authorNbh}
\end{figure}

Experiments have shown that if the number of books per author is not restricted, the final book RV will be saturated by the authors with more books, leading to worse predictions. This led us to experiment with several book limits. The evolution of results is depicted in figure~\ref{fig:authorNbh}. As shown in the graphic, predictions improve when the book limit varies from 1 to 4 and decreases after 4 for both aggregation functions. The maximum number of books per author, for the LitRec data set, is 4. From here on we will a maximum of 4 books per author.

\subsection{Aggregating book ranks}
Both book RV obtained in the prediction step are finally aggregated. The Aggregation Function consolidates the book RVs into one single vector using the WAMscore (Equation~\ref{eq:wam}). Then, we sort the final book vector, placing the most similar books at the top of the list. Finally, we select the top-n books with higher ranks, producing the top-n book list for the AU.

\begin{figure}
\includegraphics[scale=0.30]{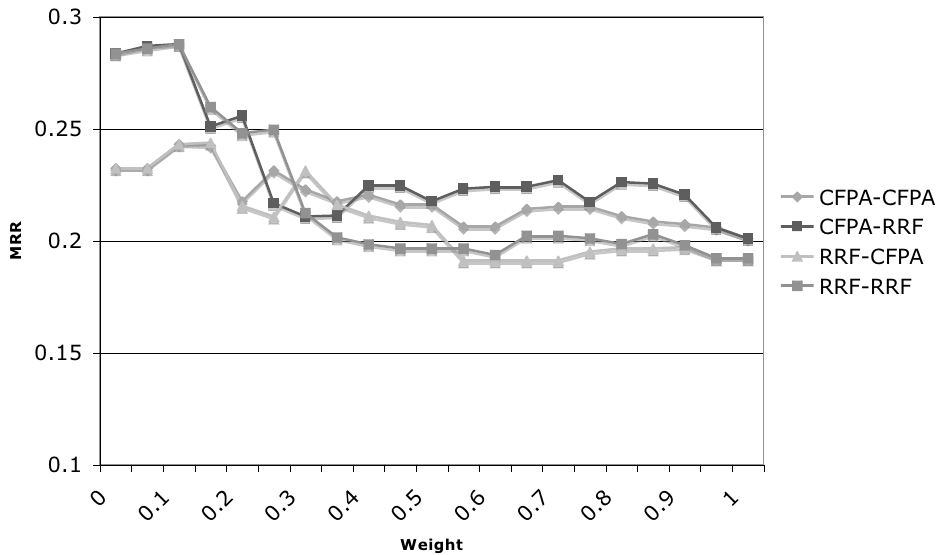}
\caption{Hybrid book recommendation algorithm MRR for all combinations of author-book score aggregation. In the legend, the first aggregation score corresponds to the author and the second corresponds to the book.}
\label{fig:finaleval}
\end{figure}

We varied the $\alpha$ parameter between $0$ and $1$ in order to assess the importance of the author in final recommendations and if final recommendations can be improved using the author. The evolution of results is shown in figure~\ref{fig:finaleval}.

As shown in the graphic, the algorithm yields the best predictions when ranks have the combination of 10\% author and 90\% book. This means that the book information is much more important than the author information to predict the books that the user will like, but the author can contribute to improve the prediction.

The graphic also outlines the evolution of all combinations of score aggregations. As expected from results obtained in the previous experiments, when author predictions use CFPAscore and books predictions use RRFscore, overall results are better. However, at the 10\%-90\% author-book combination the RRFscore-RRFscore combination can achieve the same results.

\section{Conclusions \& Future work}
In this paper we describe a hybrid book recommendation algorithm. The HBR combines two ICF algorithms that predict the books and authors the user likes. Author predictions are expanded in to a book list that is subsequently aggregated with the former book list. Finally, the resulting book list is sorted to yield the top-n book recommendations.

The HBR was tested in the LitRec data set. LitRec data set has properties and limitations that can be found in a public library. This makes it suitable to study library scenarios and  work out solutions that can be later adapted on a real public library.
% 1st question: For LitRec, ICF performance is poor.

The first of the initial goals was testing if ICF is suitable to predict books that the user will like in the LitRec conditions. Experiments led us to conclude that, the common ICF approaches yield poor predictions. When the algorithm uses co-occurrence matrices the first interesting books are placed near the third position in the book top-n.
% 2nd question: Author information can improve book recommendation results. But the book information (90\%) is more important and discriminative.
The second goal was to assess if book prediction by author selection can be used to improve overall predictions. Experiments in LitRec have shown that overall predictions can be improved using author prediction. However, a maximum number of books per author must be established, otherwise authors with more books will suffocate less productive authors, yielding one-author book top-n predictions. The maximum number for LitRec was set to 4.

We also observed that weighting the output of the both ICF algorithms differently achieve better predictions. For the LitRec data set, the contribution of choosing books by author must be smaller than book co-occurrences, i.e, book popularity. These will require experiments with a more extended combination of weights for further study.

This paper describes exploratory work in LitRec data set that open a path for further research. We intend to continue exploring LitRec. We will try to assess if book choice is related to content, user location, and the month in the book read date. Finally, the use of feature augmentation and dimensionality reduction techniques like singular value decomposition or principal component analysis will also be considered.

\bibliographystyle{abbrv}
\bibliography{sigproc}  % sigproc.bib is the name of the Bibliography in this case
\end{document}